\documentclass[aip,apl,amsmath,amssymb,reprint]{revtex4-1}

\usepackage{graphicx}
\usepackage{dcolumn}
\usepackage{bm}

\usepackage[utf8]{inputenc}
\usepackage[T1]{fontenc}
\usepackage{mathptmx}

\begin{document}

\title{Efficient demultiplexed single-photon source with a quantum dot coupled to a nanophotonic waveguide}

\author{Thomas Hummel}

\author{Claud\'{e}ric Ouellet-Plamondon}

\author{Ela Ugur}
\affiliation{Center for Hybrid Quantum Networks (Hy-Q), Niels Bohr Institute, University of Copenhagen, Blegdamsvej 17, 2100-DK Copenhagen, Denmark}

\author{Irina Kulkova}

\author{Toke Lund-Hansen}
\affiliation{Sparrow Quantum, Blegdamsvej 17, 2100-DK Copenhagen, Denmark}

\author{Matthew A. Broome}
\altaffiliation[Present address: ]{Department of Physics, University of Warwick, Gibbet Hill Road, Coventry, CV4 7AL, United Kingdom}

\author{Ravitej Uppu}
\email{ravitej.uppu@nbi.ku.dk}

\author{Peter Lodahl}
\email{lodahl@nbi.ku.dk}
\affiliation{Center for Hybrid Quantum Networks (Hy-Q), Niels Bohr Institute, University of Copenhagen, Blegdamsvej 17, 2100-DK Copenhagen, Denmark}

\date{\today}

\begin{abstract}
Planar nanostructures allow near-ideal extraction of emission from a quantum emitter embedded within, thereby realizing deterministic single-photon sources.
Such a source can be transformed into $M$ single-photon sources by implementing active temporal-to-spatial mode demultiplexing. 
We report on the realization of such a demultiplexed source based on a quantum dot embedded in
a nanophotonic waveguide. 
Efficient outcoupling ($>60\%$) from the waveguide into a single mode optical fiber is obtained with high-efficiency grating couplers.
As a proof-of-concept, active demultiplexing into $M=4$ spatial channels is demonstrated by the use of electro-optic modulators with an end-to-end efficiency of $>81\%$ into single-mode fibers.
Overall we demonstrate four-photon coincidence rates of $> 1$ Hz even under non-resonant excitation of the quantum dot. 
The main limitation of the current source is the residual population of other exciton transitions that corresponds to a finite preparation efficiency of the desired transition. 
We quantitatively extract a preparation efficiency of $15\%$ using the second-order correlation function measurements.
The experiment highlights the applicability of planar nanostructures as efficient multiphoton sources through temporal-to-spatial demultiplexing and lays out a clear path way of how to scale up towards demonstrating quantum advantages with the quantum dot sources.
\end{abstract}

\maketitle
The recent advances in experimental quantum-information processing \cite{Flamini2019,Rudolph2015,Carolan2014,Walmsley2015,Kok2007} and cryptography \cite{Carlos2018,Thomas2018,Pan2018} highlight the necessity for efficient single-photon sources.
Photons are robust carriers of quantum information and enable scalable quantum simulations.\cite{Aspuru2012,Sparrow2018,Spring2012}
These applications require efficient deterministic sources of multiple indistinguishable single photons.\cite{Sangouard2012,Motes2014,Lodahl2017}
The traditional approach to  multiphoton generation is based on probabilistic parametric downconversion or four wave mixing sources.\cite{Kaneda2018,Xiong2016,Rudolph2016}
The scaling up of the number of generated photons using such sources is limited by the low generation efficiency and the large amount of resources (detectors and optical switches) needed for heralding the photons.
Over the last decade, fundamental and technological progress in the growth and control of semiconductor quantum dots has resulted in their applicability as near-ideal single-photon emitters.\cite{Kuhlmann2015,Lodahl2015,Aharonovich2016}
Crucially, enhancing light-matter interaction through the fabrication of on-chip nanophotonic structures containing quantum dots has resulted in efficient deterministic and coherent single-photon sources.\cite{Arcari2014,Huber2015,Ding2016,Senellart2017,Lee2018}
However, the inhomogeneous broadening of the quantum dots poses a challenge in creating multiple identical sources.

An alternative route towards high-brightness multi-photon generation is by implementing active temporal-to-spatial mode demultiplexing of the emitted single-photon train from a single quantum dot.\cite{Wang2016,Lenzini2017}
Recent experiments achieved high degree of indistinguishability ($>90\%$) over long timescales,\cite{Wang2016a} which enabled the temporal demultiplexing of a quantum dot in a micropillar cavity.\cite{Wang2016}
Planar nanostructures offer the opportunity for near-unity and broadband coupling of quantum dot emission to a single propagating mode, which could enable integration of functionalities on-chip for ultimate system efficiency.
Furthermore, the polarization of the propagating mode can be engineered in the planar nanostructures, which ensures suppression of emission into unwanted polarization states.\cite{Arcari2014}

In this work, we realize temporal-to-spatial mode demultiplexing of an efficient single-photon source based on a quantum dot embedded in a nanophotonic waveguide.
We analyze the preparation efficiency of the quantum dot source and demonstrate high collection efficiency of the single-photon emission into a single mode optical fiber.
Subsequently, we perform active switching of the train of single photons and measure four-photon generation rate of $>1$ Hz of the multiphoton source.

\begin{figure}
\includegraphics[width = \columnwidth]{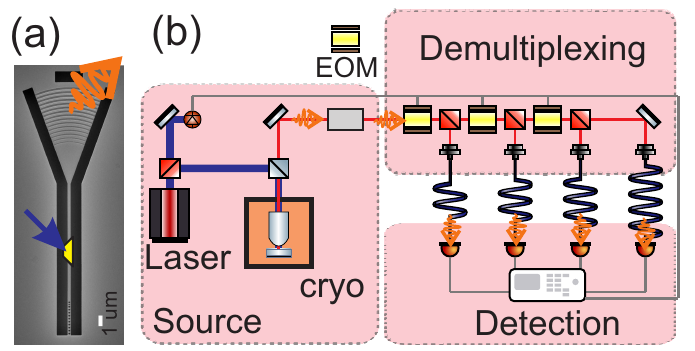}
\caption{\label{fig:Setup}
Overview of the demultiplexed four-photon source. (a) Scanning electron microscope image of the planar nanophotonic waveguide terminated with a mirror on one end and a grating outcoupler on the other.
The quantum dot embedded within the waveguide is optically excited from the top.
(b) The schematic of the experimental setup illustrates the generation, the demultiplexing and the detection of photons.
A pulsed laser excites the sample placed in a cryostat and the emitted single photons are collected through a bandpass filter and directed to the demultiplexing setup.
The four spatial-mode demultiplexer is composed of three switches in sequence, each made of an electro-optic modulator (EOM) and a polarizing beam splitter.
The switches synchronously route the photon pulse train to the four delay fibers connected to single-photon detectors.
}
\end{figure}

Figure \ref{fig:Setup}(a) shows the planar device used in the generation of the single photons.
The device consists of an indium arsenide (InAs) quantum dot embedded in a gallium arsenide (GaAs) suspended nanobeam waveguide (width = $320$ nm; height = $160$ nm; length = $14.7$ $\mu$m).
The waveguide is terminated on one end with a photonic crystal mirror and on the other end with a high-efficiency grating outcoupler.\cite{Zhou2018}
The nanobeam waveguide engineers the local density of states experienced by the quantum dot, thereby selectively coupling the emission to the waveguide mode.
The grating outcoupler is optimized to direct the polarized light in the waveguide off the chip and into a single mode optical fiber.

Figure \ref{fig:Setup}(b) shows a schematic of the optical setup used for generating and routing single photons.
The setup is broadly separated into three sections: 1) generation of single photons, 2) active temporal-to-spatial demultiplexing, and 3) high-efficiency detection and analysis.
For generating single photons, the device is cooled to a temperature of 4.2 K in a liquid helium bath cryostat with optical access.
Light from a pulsed Ti:Sapphire laser ($\lambda = 853$ nm; repetition rate $F_\textrm{rep} = 76.152$ MHz; pulse width $\approx 3$ ps) is focused using a high-NA objective to excite a single quantum dot in the nanobeam waveguide.
The quantum dot emission coupled to the waveguide is collected by the same objective at the grating outcoupler.
The collected photons are coupled into a single mode fiber through a 90:10 (transmission:reflection) beamsplitter.
As the excitation of the quantum dot is non-resonant, the emission spectrum is composed of multiple lines.
A tunable bandpass filter ($\Delta \lambda = 0.3$ nm) is used to select photons from a single exciton line before injecting into the demultiplexer. 
The multiple emission lines limit the overall efficiency of the present device, which can be readily improved in a next-generation device with the implementation of electrical control of the transitions and resonant optical excitation.\cite{Lobl2017} 
The focus in the present paper is on the proof-of-concept demonstration of multiphoton demultiplexing and a quantitative assessment of the determining parameters for scaling up, which will pave the way for optimizing all parameters in a single device.

The 4-spatial mode demultiplexer comprises of three optical switches, each built using a high-transmission electrical-broadband electro-optic modulator (EOM) and a polarizing beam splitter.
The polarized single photons after the bandpass filter are routed into 4 distinct spatial modes synchronously with the laser trigger.
The EOMs have a maximum repetition rate of $F_\textrm{max} = 1$ MHz with a $27\%$ duty cycle (rise/fall time = $7$ ns).
Since $F_\textrm{max} < F_\textrm{rep}/M$, where $M = 4$ is the number of spatial modes, photons emitted from sequential excitation pulses are not switched to different modes.
Instead, we switch $N = 20$ sequential pulses per mode such that the EOM repetition rate $F_\textrm{EOM} = F_\textrm{rep}/(N\cdot M) = 952$ kHz.

Single-mode fibers at the output ports of the demultiplexer are used to collect and temporally match the routed photons.
These fibers are connected to high efficiency super-conducting nanowire single-photon detectors (SNSPDs) with a time jitter of $\approx 100$ ps.
The photon detection times and the laser trigger are recorded using a time tagger (resolution $81$ ps).

\begin{figure}
\includegraphics[width = \columnwidth]{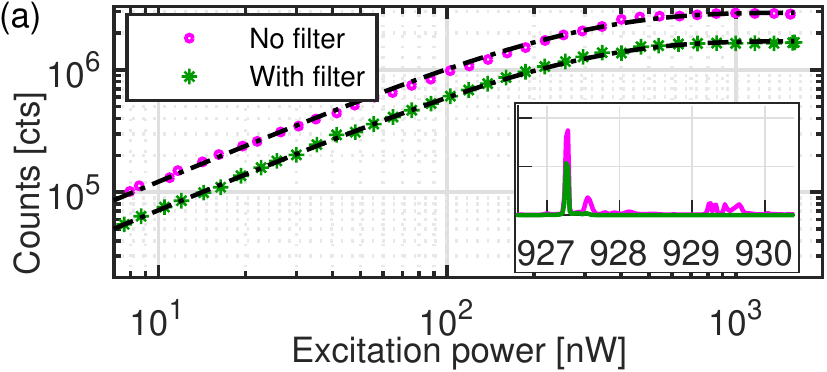}
\includegraphics[width = \columnwidth]{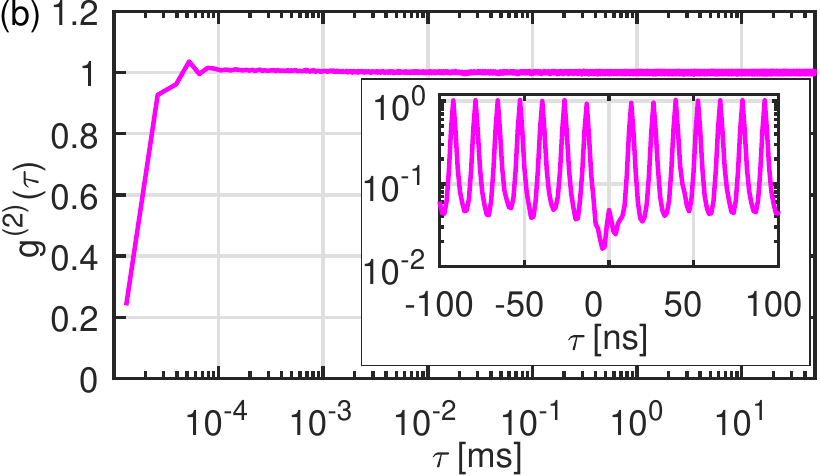}
\caption{\label{fig:Source}
(a) Measured power series of the quantum dot with ($\star$) and without (o) the narrow bandpass spectral filter.
The solid curves are fits to extract the saturation power and filter efficiency $\eta_F$.
The inset shows the emission spectra at an excitation power of $231~nW$ with and without spectral filtering.
The count rate is determined by fitting a Voight lineshape to the emission line at $927.3$ nm.
(b) Second-order correlation function $g^{(2)}(\tau)$ measured with a Hanbury Brown-Twiss setup at an excitation power of $231~nW$.
The time bin size is $1/F_{rep} = 13.1$ ns. 
$g^{(2)}(\tau)$ is normalized to the long time dynamics ($\tau = 50$ ms) and only displays very minor bunching, which indicates insignificant amount of blinking.
The inset shows short time dynamics of $g^{(2)}(\tau)$ with time bin size of $100$ ps.
Distinct peaks at $F_{rep}$ are visible with a $g^{(2)} (0) = 0.05$.
The asymmetry around $\tau = 0$ ns is an artifact of minor detector cross-talk.
}
\end{figure}
%
Before demultiplexing , the single-photon source is characterized for efficiency and purity.
First, the quantum dot emission is measured at varying excitation powers on a grating spectrometer to estimate the saturation power.
The integrated intensity of the exciton line is shown in figure \ref{fig:Source}(a) with and without the bandpass filter in collection.
The emission intensity $I_\textrm{cts}$ is modelled with a three-parameter saturation curve
\begin{equation}
    I_{\textrm{cts}}=I_0 \cdot \eta_F \cdot (1-\exp^{-{P}/{P_\textrm{sat}}}),
\label{eq:Sat}
\end{equation}
where $I_0$ is the maximum count rate, $P_\textrm{sat}$ is the saturation excitation power, and $\eta_F$ is the bandpass filter efficiency ($= 100\%$ when the filter is removed).
The quantum dot saturates at $P_\textrm{sat} \approx 236$ nW and emits $I_0 = 2.9 \times 10^6$ photons per second in the brightest transition.
The bandpass filter efficiency $\eta_F$ is measured to be $58 \%$, which results in a photon rate into the demultiplexing setup of $I_{\textrm{cts}} = 1.7 \times 10^6$ photons per second at saturation.
This photon rate corresponds to an overall emission to collection efficiency of the single-photon source of $2.3\%$, which is obtained from relating the measured count rate to the repetition rate of the excitation laser.
We define the source efficiency as the probability of delivering a photon into a single mode optical fiber according to
\begin{equation}
    \eta_S = \eta_{QD} \beta \eta_{OC} T \eta_{F},
\end{equation}
where, $\eta_{QD}$ is the efficiency of the quantum dot source, $\beta$ is the waveguide collection efficiency, $\eta_{OC}$ is the outcoupling efficiency, $T$ is the transmittivity of the collection optics, and $\eta_F$ is the bandpass filter efficiency. 
In an optimized device all sub-efficiencies could be brought close to unity, which would ultimately correspond to a deterministic source. 
The present focus is to analyze and exploit a non-optimized source for constructing a highly efficient demultiplexing setup that will lay the foundation for scaling up further.

\begin{table}
\begin{center}
\begin{tabular}{lc}
\hline
\hline
Exciton line $p_e$ & $37 \pm 1.5\%$ \\
Bright state efficiency $\eta_{b}$ & $40 \pm 4\%$  \\
Waveguide $\beta$ & $80 \pm 10 \%$ \\
Outcoupler efficiency $\eta_{OC}$ & $60 \pm 5\%$ \\
Collection optics $T$ & $69 \pm 2\%$ \\
Spectral filter $\eta_F$ & $58 \pm 2\%$ \\
\hline
\hline
\end{tabular}
\end{center}
\caption{Efficiency of the single photon source. }
\label{tableParaSource}
\end{table}

Under non-resonant excitation, the quantum dot efficiency $\eta_{QD}$ on the selected transition at saturation is limited by two processes: i) excitation of a quantum dot transition not selected by the bandpass filter or ii) coupling of the quantum dot to other states that decay radiatively or non-radiatively. 
Process i) is revealed from the emission spectra, cf.  inset of Fig. \ref{fig:Source} (a), from which the preparation efficiency $p_e$ is extracted as the fraction of photons emitted on the selected exciton state.
We measure $p_e = 37\%$ by analyzing the data with and without the bandpass filter inserted. 
Process ii) may lead to the observation of photon bunching in second-order intensity correlation function data.\cite{Davanco2014}
Figure \ref{fig:Source} (b) shows the measured long timescale dynamics of $g^{(2)}(\tau)$ at saturation using a Hanbury Brown-Twiss setup with time bin size of $ 1/F_\textrm{rep}$.
The curve was normalized to the coincidence rate at $\tau = 50$ ms.
We observe only minor bunching ($< 2\%$) implying that slow blinking to other exciton complexes plays an insignificant role. 
However, with non-resonant excitation the population of dark excitons occurs on a time scale that cannot be resolved from the $g^{(2)}(\tau)$ data. 
The dark exciton reveals in the time-resolved emission as a biexponential decay (data not shown) with the bright and the dark state decay rates being $\Gamma_{bright} = 1.3$ ns$^{-1}$ and $\Gamma_{dark} = 0.2$ ns$^{-1}$, cf. Ref. \cite{Johansen2010} for a detailed analysis of the dark exciton recombination.
Using the analysis, we estimate the bright state efficiency $\eta_b$ to be $\approx 40\%$.
The quantum dot efficiency under non-resonant excitation is thus limited to $15\%$ as $\eta_{QD} = p_e \eta_b$.
Under pulsed excitation with $F_\textrm{rep} = 76.152$ MHz, the quantum dot emits photons in the selected bright state at a rate of $11.4$ MHz. 
The measured setup efficiencies are listed in Tab. \ref{tableParaSource}, which we use to compare the expected single photon rate into the demultiplexer with the measured rate.
The planar nanostructure collects $\approx 8$MHz into the waveguide, owing to the high $\beta$-factor estimated from calculations.\cite{Thyrrestrup2018}
The transparency $T$ of the collection optics and the high-efficiency outcoupler enables a single photon rate into the single mode fiber of $3.3$ MHz.
Upon spectrally filtering the collected emission, the single photon rate in the fiber to the demultiplexer setup is $1.9$ MHz.
Employing SNSPDs with efficiency $\eta_{det} = 88\%$, we expect a photon detection rate of $1.7$ MHz, which agrees excellently with the measured single photon rate.
The single photon source purity is measured using the inset of Fig. \ref{fig:Source}(b), which shows the short timescale $g^{(2)}(\tau)$ with time bin size of $100$ ps displaying the peaks at laser repetition time of $1/F_\textrm{rep}$.
The anti-bunching at zero time delay with $g^{(2)}(0) = 0.05$ indicates high-purity single photon generation.
We note that the slight asymmetry at zero time delay is an artifact of electronic cross-talk between the two detectors.
We measured a similar value of $g^{(2)}(0)$ using avalanche photodiodes, which did not possess this electronic cross-talk.

\begin{table}
\begin{center}
\begin{tabular}{lc}
\hline
\hline
Fiber to demux $\eta_{fiber}$ & $92 \pm 2\%$\\
Switching $\eta_{sw}$ & $97 \pm 1\%$\\
Transmittance $\eta_m$ & $86 \pm 2\%$\\
Detector $\eta_{det}$ & $88\%$ \\
\hline
\hline
\end{tabular}
\end{center}
\caption{Efficiency of the four-channel demultiplexer setup.}
\label{tableParaDemux}
\end{table}

\begin{figure}
\includegraphics[width = \columnwidth]{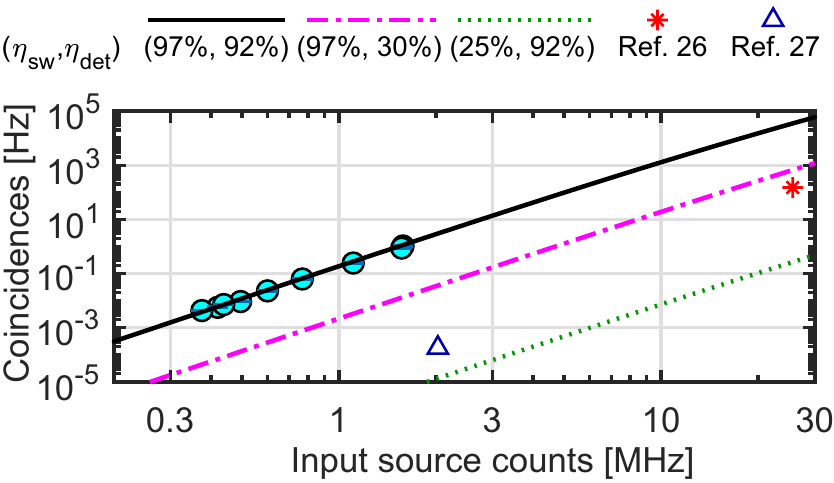}
\caption{\label{fig:Result}
Detected four-fold coincidence rate versus input source brightness (filled circles).
The expected four-fold detection rate (solid curve) is compared to a passive demultiplexer with same setup efficiency (dotted line).
A comparison of measured four-fold coincidence from Ref. \cite{Wang2016} is shown as the red star and the estimated four-fold coincidence rate from Ref. \cite{Lenzini2017} as the blue triangle.
The expected four-fold coincidence rate in our setup using avalanche photodiodes with $\eta_{det} = 30\%$ is plotted as dash dotted curve for comparison with the earlier experiments.
}
\end{figure}

The generated single photons are transmitted to the demultiplexer using a $30$ m optical fiber with transmission $\eta_{fiber} = 92\%$.
The single and four-photon-coincidence events are accumulated using the SNSPDs and the time tagger over a period of few hours.
The input single photon rate into the demultiplexer is changed by varying the excitation power.
The detected four-fold coincidence rates at different input source count rate is shown in Fig. \ref{fig:Result}, which is $>1$ Hz at saturation of the quantum dot.
The performance of the demultiplexer can modeled as follows.
With $M = 4$ spatial modes and $N = 20$ photons per mode, the probability of a single photon clicks $\rho_m^n$ in the spatial mode $m$ and the temporal mode $n$ is
\begin{equation}
\rho_m^n= \eta_{S}\eta_{m}\eta_{sw}\eta_{det} \sum^{n-1}_{k=0} \prod^{n-1}_{\varepsilon=k+1} \left( 1-\rho^\varepsilon_m\right)\rho^k_m T^{n,k}_m.
\label{eq:rho}
\end{equation}
Here, $\eta_m$ accounts for the transmission and fiber coupling efficiency in each arm of the demultiplexer and $\eta_{sw}$ is the efficiency of the switch.
We define $\rho_m^0 = 1$ and $T_m^{n,0} = 1$.
The expression in the summation takes into account the temporal response of the detector in $m$-th spatial mode and $n$-th temporal mode $T^{n,k-1}_m$ to accurately calculate the single photon detection probability for finite dead time.
We neglect the time jitter ($\approx 100$ ps) in detection as it is much smaller than the relevant timescales, $1/F_\textrm{rep}$ and $T^{n,k-1}_m$.
The transmission of spatial channels $\eta_{m}$ varies by $<9\%$ across the four channels.
The efficiency of the switch $\eta_{sw}$ determines the performance of the active demultiplexer. 
As a special limiting case a passive setup based on splitting the photons probabilistically on beam-splitters corresponds to  $\eta_{sw} = 1/M = 25\%$, while $\eta_{sw}=100\%$ would be the ideal case of a loss-less switch.

The four-fold coincidence detection rate $F_{4F}$ and the input source rate $F_{in}$ can be calculated using the efficiencies shown in Table \ref{tableParaDemux} using the following relation
\begin{eqnarray}
F_{4F} &=& \frac{F_{rep}}{M \cdot N}\sum^N_{n=1}\prod^M_{m=1}\rho^n_m,\\
F_{in} &=& F_{rep}\eta_{S}.
\label{eq:4F}
\end{eqnarray}
The expected $F_{4F}$ at any given $F_{in}$ is shown in Fig. \ref{fig:Result} for the passive and our active implementation ($\eta_{sw} = 97 \%$).
The detection efficiency and the transmittance of the demultiplexer setup is the same in both the cases.
The high efficiency of the switch in our setup allows nearly six orders of magnitude improvement over the passive demultiplexer.
The measured four-fold demultiplexed rate of the quantum dot in a nanophotonic waveguide is excellently described by our fit-parameter-free model, as shown in the figure.

In the present experiment the single-photon source had a rate of $\approx 1.7$ MHz at the input of the demultiplexer. 
By implementing electrical gates in a diode-like heterostructure the exciton charge state can be fully stabilized \cite{Lobl2017} and furthermore resonant excitation may be used to avoid excitation of residual exciton states. 
With resonant excitation, the source efficiency can be improved to $> 30 \%$, \cite{Wang2016,Senellart2017} which would lead to an expected rate of $F_{4F} \approx 38$ kHz.

A similar demultiplexing technique has been employed for multiple photon generation using free-space \cite{Wang2016} as well as chip-based switches \cite{Lenzini2017}.
These experiments employed quantum dots embedded in micropillar cavities, which do not select the polarization state of the emitted photons.
The results from these experiments are shown in Fig. \ref{fig:Result} for comparison.
These experiments employed avalanche photodiodes ($\eta_{det} = 30 \%$) for detection.
The dash dotted curve in the figure shows the expected $F_{4F}$ with the same switching efficiency as our setup, but with reduced detection efficiency.
The earlier measurements are below the expected count rates indicating a higher performance of our demultiplexing setup.
Resonant excitation in Ref. \cite{Wang2016} allowed $F_{in} > 25$ MHz, resulting in $151$ Hz of detected 4-fold coincidence rate.
Similarly, the on-chip lithium niobate switches employed in Ref. \cite{Lenzini2017} had $\eta_m < 10\%$, which severely limited the performance of the demultiplexer led to an estimated 4-fold detection rate of $0.18$ mHz.

In summary, we demonstrate highly efficient generation of polarized multiple photons through temporal-to-spatial demultiplexing of a quantum dot in a planar nanostructure.
Our proof-of-principle demonstration paves a path for scaling up quantum dot single photon sources towards experiments that reveal and exploit the quantum advantage for quantum-information processing and simulation.\cite{Flamini2019,Dalzell2018,Boixo2018}
We estimate the N-fold coincidence rates for an active demultiplexing setup, which accurately models the experiment.
The observed multi-photon generation rate is primarily limited by the quantum dot preparation efficiency that could be readily improved in next-generation samples.
A future direction may be to integrate the switches directly on-chip together with the quantum dot source.
The recent demonstration of low-loss nanomechanical switching lays out a promising route for such an integration \cite{papon2018}, which will greatly reduce the footprint required for scaling up the multi-photon source.

\begin{acknowledgments}
The authors gratefully acknowledge R\"{u}diger Schott, Andreas D. Wieck, and Arne Ludwig for growing the GaAs wafers with quantum dots and Tommaso Pregnolato for assistance in fabrication. We gratefuly acknowledge financial support from the Danish National Research Foundation (Center of Excellence “Hy-Q”), the Europe Research Council (ERC Advanced Grant “SCALE”), Horizon2020 (Marie-Sklowdowska Curie Individual Fellowship), Innovation Fund Denmark (Quantum Innovation Center “Qubiz”), and the Danish Research Infrastructure Grant (QUANTECH).
\end{acknowledgments}

\bibliography{DemuxRefs}
\end{document}